\documentstyle[twocolumn,epsf]{jpsj}
\def\runtitle{}
\def\runauthor{}

\title{
Spin Liquid State around a Doped Hole in Insulating Cuprates
}

\author
{
Takami {\sc Tohyama}\footnote{E-mail address: tohyama@imr.tohoku.ac.jp},
Yasumasa {\sc Shibata}, Sadamichi {\sc Maekawa},
Z.-X. {\sc Shen}$^1$,\\
Naoto {\sc Nagaosa}$^2$ and Lance L. {\sc Miller}$^3$
}

\inst
{
Institute for Materials Research, Tohoku University,
 Sendai 980-8577, Japan\\
$^1$Department of Applied Physics and Stanford Synchrotron
Radiation Laboratory, Stanford University,\\
Stanford, California 94305, USA\\
$^2$Department of Applied Physics, University of Tokyo,
Bunkyo-ku, Tokyo 113-8656, Japan\\
$^3$Ames Laboratory, Iowa State University, Ames, Iowa 50011, USA

}

\recdate
{August 16, 1999}

\abst
{
The numerically exact diagonalization study on small clusters of
the $t$-$J$ model with second- and third-neighbor hopping
terms shows that a novel spin liquid state is realized around
a doped hole with momentum {\bf k}=($\pi$,0) and energy $\sim$2$J$
compared with that with ($\pi$/2,$\pi$/2) in insulating cuprates,
where the spin and charge degrees of freedom are approximately
decoupled.  Our finding implies that the excitations in the
insulating cuprates are mapped onto the $d$-wave resonating
valence bond state.
}

\kword
{
spin liquid, spin-charge separation, insulating cuprates, $t$-$J$ model
}

\begin{document}
\sloppy
\maketitle

The pairing symmetry of high-$T_c$ superconductors has been
established to be of $d_{x^2-y^2}$ type, whose gap is maximum
in the direction of {\bf k}=($\pi$,0).  Such a $d$-wave gap has
been observed not only in the superconducting state but also
in the normal state for the underdoped superconductors by
angle-resolved photoemission spectroscopy (ARPES)
experiments~\cite{Loeser,Ding,Marshall}.  The normal-state gap
is called a pseudogap.  In the overdoped cuprates, the pseudogap
almost disappears and a flat band with sharp spectral
weight emerges around the ($\pi$,0) point.  Understanding of
the electronic states at ($\pi$,0) is, therefore, critical
in the field of cuprate superconductivity.

In the insulating cuprates, ARPES experiments~\cite{Wells,Kim}
have revealed the characteristic excitations below
the charge transfer gap with minimum binding energy
at {\bf k}=($\pi$/2,$\pi$/2):
While the spectrum near ($\pi$/2,$\pi$/2) consists of a sharp
peak, the ($\pi$,0) spectrum is very broad.  From the numerically
exact diagonalization study for a $t$-$J$ model with second-
and third-neighbor hoppings $t'$ and $t''$, respectively, i.e., the
$t$-$t'$-$t''$-$J$ model~\cite{Kim}, it has been found that
the quasi-particle (QP) weight at ($\pi$,0) is remarkably
reduced through weakening of the antiferromagnetic (AF) spin
correlation induced by $t'$ and $t''$.  The resulting broad
($\pi$,0) spectrum in the insulator continuously evolves into
the broad one at underdoping to the sharp one at
overdoping~\cite{Kim}.  Although the numerical results have
demonstrated a reasonable agreement with ARPES data, the physical
origin of the broadness of the ($\pi$,0) spectrum remains
to be elucidated.  Very recently, Ronning
{\it et al.}~\cite{Ronning} pursued this issue,
performing ARPES experiments on Ca$_2$CuO$_2$Cl$_2$, a parent
compound of high-$T_c$ superconductors.  The experimental data
show a $d$-wave-like dispersion along the ($\pi$/2,$\pi$/2)-($\pi$,0)
line, implying that the pseudogap is remnant of the $d$-wave
gap in the insulator.

In this Letter, we examine the dynamics of a hole in the insulator
and the spin configuration around it.  With realistic values
of $t'$ and $t''$, we find a dispersion consistent with that
predicted by the resonating valence bond (RVB)
theory~\cite{Anderson} with $d$-wave
pairing~\cite{Affleck,Laughlin,Wen}, where the incoherent motion of
the hole plays a crucial role.  These $t'$ and $t''$ also lead to
a novel spin liquid state around a hole with {\bf k}= ($\pi$,0)
and excitation energy of $\sim$2$J$,
where the hole causes an antiphase in the AF spin background.
Correspondingly, the spin and charge degrees of freedom are
approximately decoupled there, as seen from the spin and charge
correlation functions.  In contrast to the ($\pi$,0) state,
the low-energy states near ($\pi$/2,$\pi$/2) exhibit
AF behavior.  The implication of the spin liquid state
is discussed in connection with the $d$-wave RVB state as well
as the pseudogap.

We first clarify the importance of the incoherent motion of
a hole by comparing the calculations of the single-hole dispersion
based on (i) a self-consistent Born approximation
(SCBA)~\cite{Schmitt,Kane,Martinez} for the $t$-$t'$-$t''$-$J$ model
and (ii) a spin-density wave (SDW) mean-field approximation for
a Hubbard model containing the long-range hopping terms
($t$-$t'$-$t''$ Hubbard).  In SCBA, the self-energy of a hole is
determined by taking into account the coupling to the spin
background.  This scheme explains consistently the reduction of
the dispersion from $t$ to $J$ and the incoherent part of the
spectral function extending over the energy $\sim t$.
The SCBA results are in good agreement with those of the
numerically exact diagonalization (ED) method~\cite{Xiang}.
In contrast to SCBA, the SDW approximation gives only a coherent
motion of a hole.

The $t$-$t'$-$t''$-$J$ model is given by
\begin{eqnarray}
H&=& J\sum\limits_{\left<i,j\right>_{1{\rm st}}}
      {{\bf S}_i}\cdot {\bf S}_j
    -t\sum\limits_{\left<i,j\right>_{1{\rm st}} \sigma }
    c_{i\sigma }^\dagger c_{j\sigma } \nonumber \\
&& {} -t'\sum\limits_{\left<i,j\right>_{2{\rm nd}} \sigma }
    c_{i\sigma }^\dagger c_{j\sigma }
     -t''\sum\limits_{\left<i,j\right>_{3{\rm rd}} \sigma }
    c_{i\sigma }^\dagger c_{j\sigma }+{\rm H.c.}\;,
\label{H}
\end{eqnarray}
where the summations $\left< i,j \right>_{1{\rm st}}$,
$\left< i,j \right>_{2{\rm nd}}$ and $\left< i,j \right>_{3{\rm rd}}$
run over first-, second- and third-nearest-neighbor pairs, respectively.
No double occupancy is allowed, and the rest of the notation is
standard.  From the previous study~\cite{Kim}, it has been known that
$t$=0.35~eV, $t'$=$-$0.12~eV, and $t''$=0.08~eV for Sr$_2$CuO$_2$Cl$_2$
and Bi$_2$Sr$_2$CaCu$_2$O$_{8+\delta}$.  These values reveal
consistent results of the Fermi surface topology with experimental data.
The SDW dispersion in the AF state of the $t$-$t'$-$t''$ Hubbard model
is given by~\cite{Duffy}
\begin{eqnarray}
E^{\pm}_{\bf k}=\varepsilon_{\bf k}\pm E^0_{\bf k}-\mu \ ,
\label{SDW1}
\end{eqnarray}
with
\begin{eqnarray}
\varepsilon_{\bf k}&=&-4t'\cos k_x\cos k_y-2t''\left( \cos2k_x +
\cos 2k_y \right) \ ,\\
E^0_{\bf k}&=&\sqrt{4t^2\left( \cos k_x + \cos k_y \right)^2 +
\left(Um\right)^2}.
\label{SDW2}
\end{eqnarray}
The chemical potential $\mu$ and the magnetization $m$ are determined
by solving a set of self-consistent equations.  We use the same
values of $t$, $t'$ and $t''$  as those for the $t$-$t'$-$t''$-$J$
model, and $U$/$t$=10 ($U$ is the on-site Coulomb interaction).
\begin{figure}[t]
\vspace{5mm}
\epsfxsize=8.0cm
\centerline{\epsffile{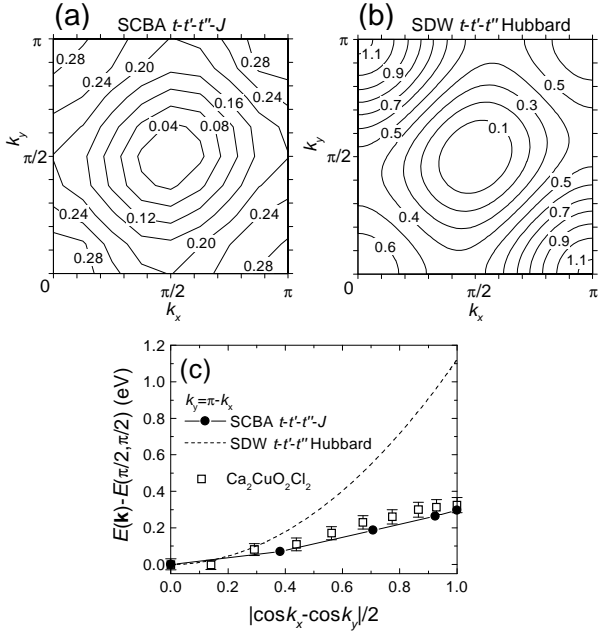}}
\caption{(a) Contour plot of SCBA single-hole dispersion of the
$t$-$t'$-$t''$-$J$ model on a 16$\times$16 lattice with $t$=0.35~eV,
$t'$=$-$0.12~eV, $t''$=0.08~eV and $J$/$t$=0.4.  The numbers in
the plot are energy values in units of eV relative to the
($\pi$/2,$\pi$/2) point.  (b) The same as (a) but for SDW
mean-field approximation results of the half-filled $t$-$t'$-$t''$ Hubbard
model with the same hopping parameters and $U$/$t$=10.
(c) $d$-wave plot of the dispersions along ($\pi$/2,$\pi$/2) to
($\pi$,0).  Dashed line: SDW result of the $t$-$t'$-$t''$ Hubbard
model.   Solid circles: SCBA results of the $t$-$t'$-$t''$-$J$
model.  Open squares: ARPES data of Ca$_2$CuO$_2$Cl$_2$ taken
from ref.~6.}
\label{fig:1}
\end{figure}

Figures~1(a) and 1(b) show the contour plots of the single-hole
QP dispersion obtained by the SCBA for the $t$-$t'$-$t''$-$J$ model
on a 16$\times$16 lattice and by the SDW approximation for the
$t$-$t'$-$t''$ Hubbard model, respectively.  The SCBA contour is
isotropic and thus consistent with the ARPES dispersion in
Ca$_2$CuO$_2$Cl$_2$ that is isotropic around ($\pi$/2,$\pi$/2),
while the SDW one is not.  In addition, the width of the SDW dispersion
is remarkably wider than that for the SCBA results.  This is
clearly seen when we plot the dispersion along ($\pi$/2,$\pi$/2) to
($\pi$,0).  In Fig.~1(c), the energies relative to the
($\pi$/2,$\pi$/2) point, $E({\bf k})-E(\pi/2,\pi/2)$, are plotted
against $\left|\cos k_x-\cos k_y\right|/2$, together with the data
from Ca$_2$CuO$_2$Cl$_2$ which are well fitted by a straight
line~\cite{Ronning}.  What is the origin and implication of the
difference between the two results?  Figure~2(a)
shows the dependence of the energy difference
$E(\pi,0)-E(\pi/2,\pi/2)$ on the values of $t'$ and $t''$.
The variable $\alpha$ represents a scaling factor of $t'$ and
$t''$ as $t'=\alpha t'_0$ and $t''=\alpha t''_0$, being
$t'_0$=$-$0.12~eV and $t''_0$=0.08~eV.  The SDW result is given
by a function $4\alpha\left(2t''_0-t'_0\right)$ [see Eq.~(2)],
independent of $U$.  The result of the $t$-$t'$-$t''$-$J$ model
deviates remarkably from the SDW line and has a tendency to
saturate around $\alpha\sim 1$.  In addition to the slow
increase of the energy with increasing $\alpha$, the wave-function
renormalization $Z$ (ref.~16) at ($\pi$,0) is dramatically
suppressed from $Z$=0.87 ($\alpha$=0) to 0.24 ($\alpha$=1).
This is again in contrast to the SDW result, where $Z$ is
always one.  At $\alpha\sim 1$, the energy difference is approximately
0.8$t$=2$J$.  This number is of crucial importance because
the incoherent spectrum of a hole starts from the energy position
of about 2$J$ above the spectrum of QP at ($\pi$/2,$\pi$/2)
(see insets in Fig.~4).  Although the energy of the QP
spectrum at ($\pi$,0) is almost the same as that at
($\pi$/2,$\pi$/2) in the $t$-$J$ model, the former increases
with increasing $\alpha$ and the spectrum merges into the
incoherent spectra for $\alpha$$\sim$1.  As a result,
the difference between $E$($\pi$,0) and $E$($\pi$/2,$\pi$/2)
saturates and $Z$ becomes small.
\begin{figure}[t]
\vspace{5mm}
\epsfxsize=8.0cm
\centerline{\epsffile{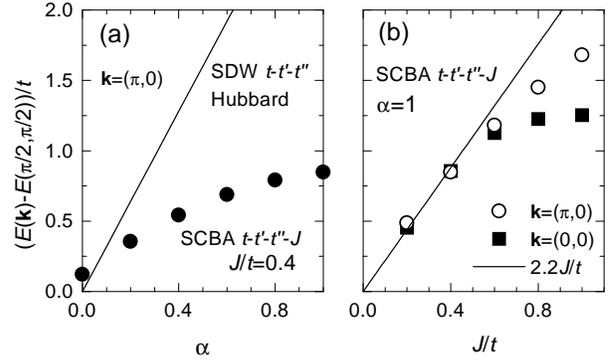}}
\caption{(a) The dependence of the energy difference
$E(\pi,0)-E(\pi/2,\pi/2)$ (in units of $t$=0.35~eV) on the values
of $t'$ and $t''$.  The variable $\alpha$ represents a scaling
factor of $t'$ and $t''$ as $t'(t'')=\alpha t'_0(t''_0)$, being
$t'_0$=$-$0.12~eV and $t''$=0.08~eV.  Solid line: SDW result of
the $t$-$t'$-$t''$ Hubbard model given by
$4\alpha\left(2t''_0-t'_0\right)$.   Solid circles: SCBA results
of the $t$-$t'$-$t''$-$J$ model on a 16$\times$16 lattice with
$J$/$t$=0.4.  (b) $\left(E({\bf k})-E(\pi/2,\pi/2)\right)$
versus $J$/$t$ in the case of $\alpha$=1.  The solid line
represents 2.2$J$/$t$ known as the band width from
($\pi$/2,$\pi$/2) to (0,0).}
\label{fig:2}
\end{figure}

The above results indicate that when realistic values of $t'$
and $t''$ ($\alpha$=1) are employed in the $t$-$t'$-$t''$-$J$
model, the dispersion from ($\pi$/2,$\pi$/2) to ($\pi$,0) is
sensitive to the value of $J$.  We now plot, in Fig.~2(b),
the energy difference $E(\pi,0)-E(\pi/2,\pi/2)$ as a function
of $J$/$t$, together with the difference between $E(0,0)$ and
$E(\pi/2,\pi/2)$ that has been known to be scaled as
$\sim$2.2$J$~\cite{Wells}.  We find that $E(\pi,0)-E(\pi/2,\pi/2)$
has the same $J$ dependence as $E(0,0)-E(\pi/2,\pi/2)$, at least,
in the realistic range $J$/$t$=0.2$\sim$0.6.  Therefore, the
dispersion of the hole is governed by the spin degree of freedom.
This explains why the dispersion is isotropic around
($\pi$/2,$\pi$/2).  Such an isotropic dispersion whose width
is controlled by $J$ is similar to that from the flux phase or
$d$-wave RVB picture~\cite{Affleck,Laughlin,Wen}.
\begin{figure}[t]
\vspace{5mm}
\epsfxsize=8.0cm
\centerline{\epsffile{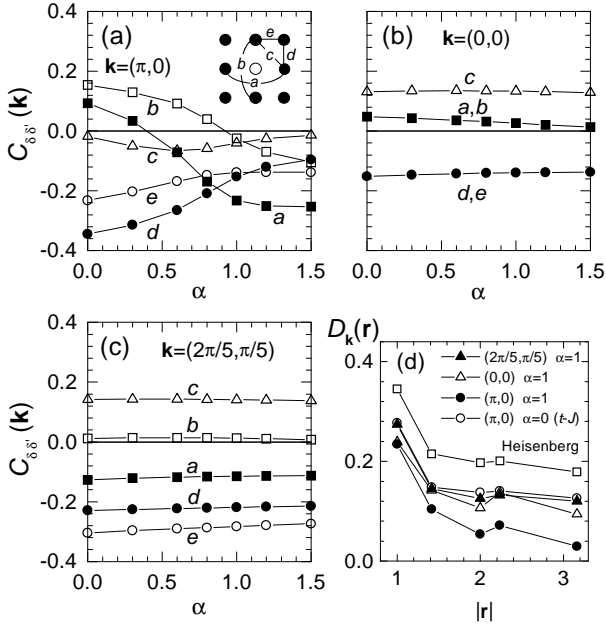}}
\caption{(a) Spin correlation $C_{\delta \delta'}({\bf k})$ of the
$t$-$t'$-$t''$-$J$ model on a 20-site cluster with one hole.
$t$=0.35~eV and $J$/$t$=0.4.  The variable $\alpha$ represents
a scaling factor of $t'$ and $t''$ as $t'(t'')=\alpha t'_0(t''_0)$,
being $t'_0$=$-$0.12~eV and $t''$=0.08~eV.  {\bf k}=($\pi$,0).
(b) The same as (a) but for {\bf k}=(0,0).  (c) The same as (a)
but for {\bf k}=(2$\pi$/5,$\pi$/5).  The labels in (a), (b) and (c)
denote configurations around a hole shown in the inset of (a).
(d) The staggered spin correlation $D_{\bf k}({\bf r})$ at several
momenta for $\alpha$=0 and 1 of the 20-site $t$-$t'$-$t''$-$J$
cluster as a function of $|{\bf r}|$.  The result for the Heisenberg
model is also shown for comparison.
}
\label{fig:3}
\end{figure}

We now examine the spin and charge characteristics of the QP state
with {\bf k}=($\pi$,0).  Figure~3(a) shows the spin correlation
around a hole defined as $C_{\delta,\delta'}\left({\bf k}\right)
\equiv \sum_i \left< \phi_{\rm QP}({\bf k} ) \left| n_i^h
{\bf S}_{i+\delta} \cdot {\bf S}_{i+\delta'} \right|
\phi_{\rm QP}({\bf k})\right>$, as a function of $\alpha$.
Here, $\delta$ and $\delta'$ denote two sites around the hole
following the labeling convention shown in the inset.
$n_i^h$ is the hole-number operator at site $i$, and
$\phi_{\rm QP}({\bf k})$ represents the wave function of
the QP state with momentum {\bf k}.  The numerically exact
diagonalization method is used for a 20-site cluster.
For $\alpha$=0, i.e., the $t$-$J$ model, the spin correlations
in the $a$ and $b$ configurations [see the inset of Fig.~3(a)]
are positive, i.e., ferromagnetic (FM), while those for $d$ and
$e$ are AF~\cite{Inoue}.  With increasing $\alpha$,
the correlation of $a$ changes from FM to AF and saturates near
$\alpha\sim 1$.  The $b$ configuration shows a similar but
slow change.  The configuration $d$ exhibits the opposite behavior,
reducing AF correlation with increasing $\alpha$, which again
nearly saturates around $\alpha$=1.  These behaviors from
$\alpha$=0 to $\sim$1 imply that the spin background of the
($\pi$,0) state changes from a N\'eel-like state to a
{\it spin liquid} state in which spins have no tendency to order.
It is interesting that the spin liquid state is accompanied by
an antiphase of spins around the hole, as evidenced by AF
correlation of the {\it a} configuration for $\alpha$$\sim$1.
Since the correlation of {\it b} is very small there,
the antiphase of spins is one-dimensional (1D).
This situation is similar to the case of a doped 1D insulator
in which the spin-charge separation occurs~\cite{Kim2}.
In contrast to the ($\pi$,0) state, QP states with different
momenta do not show such remarkable changes as a function of
$\alpha$, as shown in Figs.~3(b) and 3(c).  We note that
the AF correlation of the $a$ configuration for
{\bf k}=(2$\pi$/5,$\pi$/5) is not related to the spin
liquid state [see also Fig.~3(d)] but due to a twisted
character of the spin correlation seen in the one-hole
ground state~\cite{Inoue}.

The staggered spin correlation
$D_{\bf k}({\bf r}) \equiv \sum_i P_i({\bf r})\\
\left< \phi_{\rm QP}({\bf k})
\left| {\bf S}_{i+{\bf r}} \cdot {\bf S}_i \right|
\phi_{\rm QP}({\bf k})\right>/N_s$ also supports the concept
of the spin liquid state with ($\pi$,0).  Here, ${\bf r}$
is the vector connecting two sites, $N_s$ is the total
number of sites, and $P_i({\bf r})$=$\pm$1 depending on
whether the two sites are on the same sublattice.
Figure~3(d) shows $D_{\bf k}({\bf r})$ at several momenta
for $\alpha$=0 and 1, together with the spin correlation of
the Heisenberg model which shows AF long-range order in the
thermodynamic limit.  Except for the case of $\alpha$=1 at
($\pi$,0) (filled circles), $D_{\bf k}({\bf r})$ shows
$|{\bf r}|$ dependence similar to that of the Heisenberg
model, indicating the presence of long-range order.
The ($\pi$,0) state for $\alpha$=1, however, shows a rapid
decrease of the correlation with increasing $|{\bf r}|$.
This is again consistent with the spin liquid concept.
\begin{figure}[t]
\vspace{5mm}
\epsfxsize=8.0cm
\centerline{\epsffile{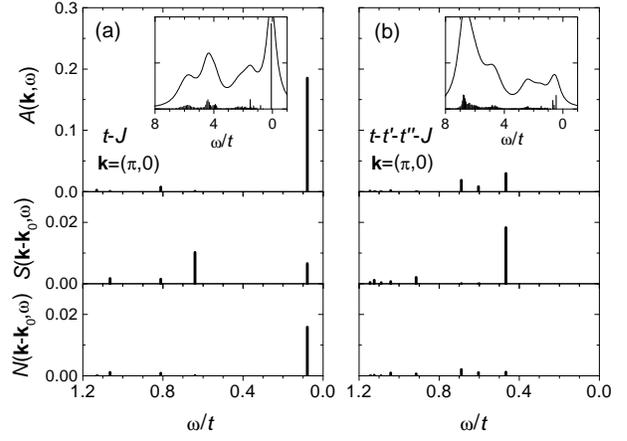}}
\caption{$A({\bf k},\omega)$ at half-filling, and
$S({\bf q},\omega)$ and $N({\bf q},\omega)$ with
{\bf q}={\bf k}-{\bf k}$_0$ at one-hole doping for a 20-site
$t$-$t'$-$t''$-$J$ model.  $J$/$t$=0.4.  {\bf k}=($\pi$,0)
and {\bf k}$_0$=(3$\pi$/5,$-\pi$/5).  The energy $\omega$
in $A({\bf k},\omega)$ is measured from the QP peak at
{\bf k}$_0$.  (a) $t$-$J$ and (b) $t$-$t'$-$t''$-$J$ with
realistic values of $t'$ and $t''$ ($\alpha$=1).
The height of the bars represents the spectral intensity.
The insets in (a) and (b) show $A({\bf k},\omega)$ with
full-energy scale.  The delta functions (vertical bars) are
broadened by a Lorentzian with a width of 0.3$t$ (solid lines).}
\label{fig:4}
\end{figure}

The dynamical properties of spin and charge degrees of freedom
also provide us with useful information on the QP state with ($\pi$,0).
We calculate the dynamical spin and charge correlation functions,
$S({\bf q},\omega)$ and $N({\bf q},\omega)$~\cite{Tohyama},
on the 20-site cluster with one hole, the Hilbert space of
which is equivalent to that for the final state of the single-hole
spectral function $A({\bf k},\omega)$.  The momentum transfer
is given by {\bf q}={\bf k}$-${\bf k}$_0$, where {\bf k}$_0$ is
the ground-state momentum of the one-hole system.  In Fig.~4,
$A({\bf k},\omega)$ with {\bf k}=($\pi$,0) is compared with
$S({\bf q},\omega)$ and $N({\bf q},\omega)$, where we take
{\bf k}$_0$=(3$\pi$/5,$-\pi$/5)~\cite{GS}.  In the $t$-$J$ model,
both spin and charge components are involved in the QP state at
$\omega$/$t$=0.08~\cite{Maekawa}, while only spin component
remains in the lowest-energy state at $\omega$/$t$=0.47 in the
$t$-$t'$-$t''$-$J$ model as seen in Fig.~4(b).  This is because
of the separation of the spin and charge degrees of freedom.
However, the separation is incomplete unlike the 1D case of the
$t$-$J$ model~\cite{Kim2}.  We note that the results of the
$t$-$t'$-$t''$-$J$ model are smoothly connected to those of
the $t$-$J$ model, showing no separation, as seen in Fig.~4(a).
Much of the weight of the charge excitation shifts to a higher
energy region where the spectral function has large weight
($\omega/t\sim$7) [see the inset of Fig.~4(b)].
For {\bf k}=(2$\pi$/5,$\pi$/5), which is close to
($\pi$/2,$\pi$/2), the QP involves both spin and charge
components independent of $\alpha$ (not shown here)~\cite{Maekawa}.

We have shown that the spin background in the ($\pi$,0) state
in the insulator behaves like a spin liquid state.  At the same
time, the spin and charge degrees of freedom are nearly separated
in the state~\cite{Martins}.  The almost linear behavior of the dispersion near
($\pi$,0) in the $d$-wave plot shown in Fig.~1(c) indicates that
the spin liquid state has a $d$-wave gap.  All of these facts
tempt us to identify the state with the $d$-wave RVB state
discussed in refs.~8$-$10.  However, we have to distinguish
the present spin liquid state with the RVB state: the former
is seen in the one-hole state with {\bf k}=($\pi$,0) and
the excitation energy of $\sim$2$J$, while the $d$-wave RVB
theory predicts a spin liquid state as well as the spin-charge
separation which are independent of the momentum of a doped hole.
In fact, the present results reveal that hole states near
($\pi$/2,$\pi$/2), i.e., the low-energy states, are not
effective to destroy the AF long-range order [see Fig.~3(d)].
However, the states near ($\pi$,0), i.e., the high-energy
($\sim 2J$) states, are in good agreement with the excited
states predicted by the $d$-wave RVB theory.  Therefore,
the spin background changes from the AF state in the low-energy
region to the spin liquid one in the high-energy region~\cite{Hsu}.
The $d$-wave RVB theory might thus be regarded as
an effective theory to describe the states in the high-energy region.

With hole doping, the energy of the ($\pi$,0) state
shifts to the Fermi level, maintaining the broadness~\cite{Kim}
that is an indication of the spin liquid state.
Since the pseudogap with the order of $J$ (high-energy pseudogap)
is determined by the position of the the broad
spectrum~\cite{Marshall}, the spin liquid concept is the most
probable explanation of the origin of the high-energy pseudogap
in underdoped cuprates~\cite{Ronning}.

In summary, we have examined the dynamics of a doped hole and
spin correlation around it in insulating cuprates by using the
$t$-$t'$-$t''$-$J$ model.  In contrast to the AF state near
($\pi$/2,$\pi$/2), we have found a novel spin liquid state
around the hole with {\bf k}=($\pi$,0) and excitation energy of
$\sim2J$, where the spin
and charge degrees of freedom are approximately decoupled.
Our findings imply that the excitations in the insulating
cuprates are mapped onto the $d$-wave RVB state.

This work was supported by the Ministry
of Education, Science, Sports and Culture of Japan, CREST and NEDO.
The numerical calculation was performed in the supercomputing
facilities in ISSP, University of Tokyo, and IMR,
Tohoku University.

\end{document}